\documentclass[aps,prx,twocolumn,showpacs,psfig,superscriptaddress,longbibliography]{revtex4-2}
\usepackage{textcomp}
\usepackage{times}
\usepackage{graphicx}
\usepackage{float}
\usepackage{latexsym,amsmath,amssymb,bm,euscript}
\usepackage{color}
\usepackage{subfigure}
\usepackage{epstopdf}
\usepackage[colorlinks=true,linkcolor=blue,citecolor=blue]{hyperref}
\usepackage{hyperref}
\usepackage{soul}
\usepackage[normalem]{ulem}
\usepackage{mathrsfs}
\usepackage{amsmath}
\usepackage{lettrine}
\usepackage{bbding}
\usepackage{xspace}
\usepackage{textcomp}

\graphicspath{{./Fig/}}

\begin{document}
\title{Significant Inverse Magnetocaloric Effect induced by Quantum Criticality}
\author{Tao Liu} 
\thanks{These authors contributed equally to this work.}
\affiliation{School of Science, Hunan University of Technology, Zhuzhou 412007, China}
\affiliation{School of Physics, Beihang University, Beijing 100191, China}

\author{Xin-Yang Liu}
\thanks{These authors contributed equally to this work.}
\affiliation{School of Physics, Beihang University, Beijing 100191, China}

\author{Yuan Gao}
\affiliation{School of Physics, Beihang University, Beijing 100191, China}

\author{Hai Jin}
\affiliation{Department of Astronomy, Tsinghua Center for Astrophysics, 
Tsinghua University, Beijing 100084, China}

\author{Jun He} 
\affiliation{School of Science, Hunan University of Technology, Zhuzhou 412007, China}

\author{Xian-Lei Sheng}
\affiliation{School of Physics, Beihang University, Beijing 100191, China}

\author{Wentao Jin} 
\affiliation{School of Physics, Beihang University, Beijing 100191, China} 

\author{\\ Ziyu Chen}
\email{chenzy@buaa.edu.cn}
\affiliation{School of Physics, Beihang University, Beijing 100191, China}

\author{Wei Li}
\email{w.li@buaa.edu.cn}
\affiliation{School of Physics, Beihang University, Beijing 100191, China}
\affiliation{International Research Institute of Multidisciplinary Science, Beihang University, Beijing 100191, China}
\affiliation{{Institute of Theoretical Physics, Chinese Academy of Sciences, Beijing 100190, China}}

\begin{abstract}
The criticality-enhanced magnetocaloric effect (MCE) near 
a field-induced quantum critical point (QCP) in the spin 
systems constitutes a very promising and highly tunable alternative 
to conventional adiabatic demagnetization refrigeration. 
Strong fluctuations in the low-$T$ quantum critical regime
can give rise to {a} large thermal entropy change and thus 
significant {cooling effect} when approaching the QCP. 
In this work, {through efficient and accurate many-body 
calculations,} we show there exists a significant inverse MCE
(iMCE) in the spin-1 quantum chain materials
(CH$_3$)$_4$NNi(NO$_2$)$_3$ (TMNIN) and 
NiCl$_2$-4SC(NH$_2$)$_2$ (DTN), where DTN
has substantial low-$T$ refrigeration capacity 
while requiring only moderate magnetic fields. 
The iMCE characteristics, including the adiabatic 
temperature change $\Delta T_{\rm ad}$, isothermal 
entropy change $\Delta S$, differential Gr\"uneisen parameter, 
and the entropy change rate, are {obtained} with {quantum} many-body 
{calculations at finite temperature}. The cooling performance, i.e., 
the efficiency factor and hold time, of the two compounds is also discussed. 
Based on the many-body calculations on realistic models for 
the spin-chain materials, we conclude that the compound DTN 
constitutes a very promising and highly efficient quantum magnetic 
coolant with pronounced iMCE properties. We advocate that such 
quantum magnets can be used in cryofree refrigeration for 
space applications and quantum computing environments.
\end{abstract}

\date{\today}
\maketitle

\section{Introduction}
The magnetocaloric effect (MCE) represents a significant adiabatic 
temperature change of a magnet as a response to the varying 
external magnetic fields~\cite{Warburg1881,Weiss1917,Tegus2002Transition,
Smith2013}. Historically, the first sub-Kelvin regime cooling was realized 
through adiabatic demagnetization refrigeration (ADR)~\cite{Lounnasmaa1974}. 
Recently, low-$T$ magnetic refrigeration gets refreshed research 
interest due to its important applications in space technology
\cite{Hagmann1995,Shirron2007} and the cryofree sub-Kelvin environment 
for quantum computers~\cite{Jahromi2019nasa}. It is of great research 
interest to pursue novel MCE refrigerants that provide
higher cooling powers and can reach lower temperatures.
Among others, the quantum spin-chain materials
with enhanced MCE characterized by the universally 
diverging Gr\"uneisen ratio near the quantum critical point 
(QCP)~\cite{Zhitomirsky2003,Zhitomirsky2004,Zhu2003,
Garst2005,Honecker2009,Sharples2014,Orend2017NENB},
has been proposed as a very promising quantum critical coolant 
for magnetic refrigeration~\cite{Wolf2011,Lang2013} 
and an excellent alternative to the conventional ADR.
 
In most magnetic materials, the spin degrees of freedom in the 
system eventually ``solidify" into a long-range order as cooled 
down to {sufficiently} low temperatures. In such magnetically 
ordered phase, the spin states are practically non-tunable by 
external fields {(due to the existence of giant Weiss molecular fields)}
and the corresponding MCE, temperature or entropy change
as a response to fields, are usually negligible. Paramagnetic salts,
like CrK(SO$_4$)$_2$$\cdot$12H$_2$O 
(Chromic Potassium Alum, CPA) and Fe(SO$_4$)$_2$NH$_4$ 
$\cdot$ 2H$_2$O (Ferric Ammonium Alum, FAA), etc, 
host nearly non-interacting spins {and do not order 
at very low temperatures}, are widely used in ADR 
as spin ``gas" refrigerant~\cite{Shirron2007}.
It is commonly believed that the spin interactions
are ``harmful" {for good ADR coolants}, as they usually lead 
to magnetic ordering {as $T$ lowers} and spoil the MCE properties.

Nevertheless, there is exotic exception to this classic
fate of interacting quantum spins. {In low dimensional 
quantum magnets~\cite{Schollwoeck2004QMag},} the 
quantum fluctuations can be strong enough to prevent 
the spins from classical ordering even at $T=0$. One prominent
case is the field-induced QCP in the correlated quantum 
magnetic materials. The enhanced quantum fluctuations 
near the QCP can significantly influence the thermodynamics 
in the quantum critical regime at low temperature~\cite{Sachdev2000}. 
When approaching the QCP, the spin system would experience 
a significant isothermal entropy change, which can be translated 
into a considerable temperature decrease under the adiabatic condition.
Such quantum criticality-enhanced MCE is reflected in a 
diverging Gr\"uneisen parameter [adiabatic temperature
change rate, cf. Eq.(\ref{Eq:Gamma}) below], $\Gamma_B \sim T^{-1/z\nu}$
under external field $B$, with $z$ and $\nu$ the dynamical and critical 
exponents related to the universality class of the QCP ~\cite{Zhu2003,Garst2005}.
Notably, such intriguing quantum critical phenomena in 
low-temperature thermodynamics also provides a sensitive 
probe of QCP {in experiments}~\cite{Gegenwart2016}.

The low-temperature thermodynamic properties of a typical 1D quantum 
magnetic system --- the spin-1/2 Heisenberg chain (HAFC) --- have been
intensively explored, where a pronounced MCE was predicted~\cite{Zhitomirsky2003,Zhitomirsky2004} 
and also observed in the compound [Cu($\mu$-C$_2$O$_4$)(4-aminopyridine)$_2$(H$_2$O)]$_n$ (CuP). 
The lowest achievable temperature with the spin-chain coolants 
has no principal limitation (as long as the inter-chain
interactions are negligible), and the high efficiency factor
as well as long hold time makes the spin-1/2 HAFC materials 
very promising quantum critical refrigerants~\cite{Wolf2011,Lang2013}. 

Nevertheless, there is still plenty of room for further 
improving the performance of quantum magnetic refrigerants. 
As reported in Ref.~\cite{Wolf2011}, for the spin-1/2 HAFC 
compound CuP, one has to start from a rather high magnetic field, 
e.g., 7~T, significantly above the critical field $B_c\simeq4.09$~T. 
In CuP, there indeed exists significant MCE in the range $B > B_c$, 
while in the smaller-field side, i.e., $B \in [0, B_c]$, 
rather weak (inverse) MCE was observed. Therefore, 
the generated strong magnetic fields (of 7~T) have actually not 
been fully exploited in the case of spin-1/2 compound CuP. 
On the other hand, popular paramagnetic refrigerants are typically 
with high spin $S\geq 1$, as larger entropy changes are generally 
expected for higher spin systems. For example, 
CPA and FAA are with $S=3/2$ and $5/2$, respectively, 
and the Gadolinium Gallium Garnet (GGG) is even with $S=7/2$.  
However, there is only a few studies on quantum spin models~\cite{Honecker2009} 
and materials~\cite{Orend2017NENB} with spin higher than $S=1/2$.

In this work, we systematically investigate the iMCE properties 
of the $S=1$ quantum chain models and materials {with many-body 
simulations}. Special emphases are put on two typical spin-1 
HAFC materials, i.e., (CH$_3$)$_4$NNi(NO$_2$)$_3$ (TMNIN) 
and NiCl$_2$-4SC(NH$_2$)$_2$ (DTN), whose magneto-thermodynamic 
properties can be accurately simulated by the thermal-state linearized 
tensor renormalization group (LTRG) approaches~\cite{Li2011,Dong2017}. 
In particular, we find the compound DTN has iMCE refrigeration capacity 
comparable to the MCE of spin-1/2 HAFC material CuP, while requiring 
only a moderate external magnetic field of $B_c \simeq 3$~T in the 
cooling process, much smaller than the latter requires. Due to the pronounced 
cooling effects and excellent thermal transport properties~\cite{Sun2009DTN} 
in DTN, we propose DTN constitutes a very promising iMCE refrigerant 
with only a moderate magnetic field and competitive performance,
very suitable for practical refrigeration applications. 

The rest parts of the article is arranged as follows. 
We present the spin-1 chain models, their thermal 
many-body simulations, and the related spin-1 materials in 
Sec.~\ref{Sec:ModMeth}. Our main results on iMCE of the 
spin-1 chains are shown in {Sec.~\ref{Sec:iMCE}}, and
Section~\ref{Sec:Conclusion} is devoted to the conclusion and outlook.

\begin{figure}[tb]
\includegraphics[width=1\linewidth]{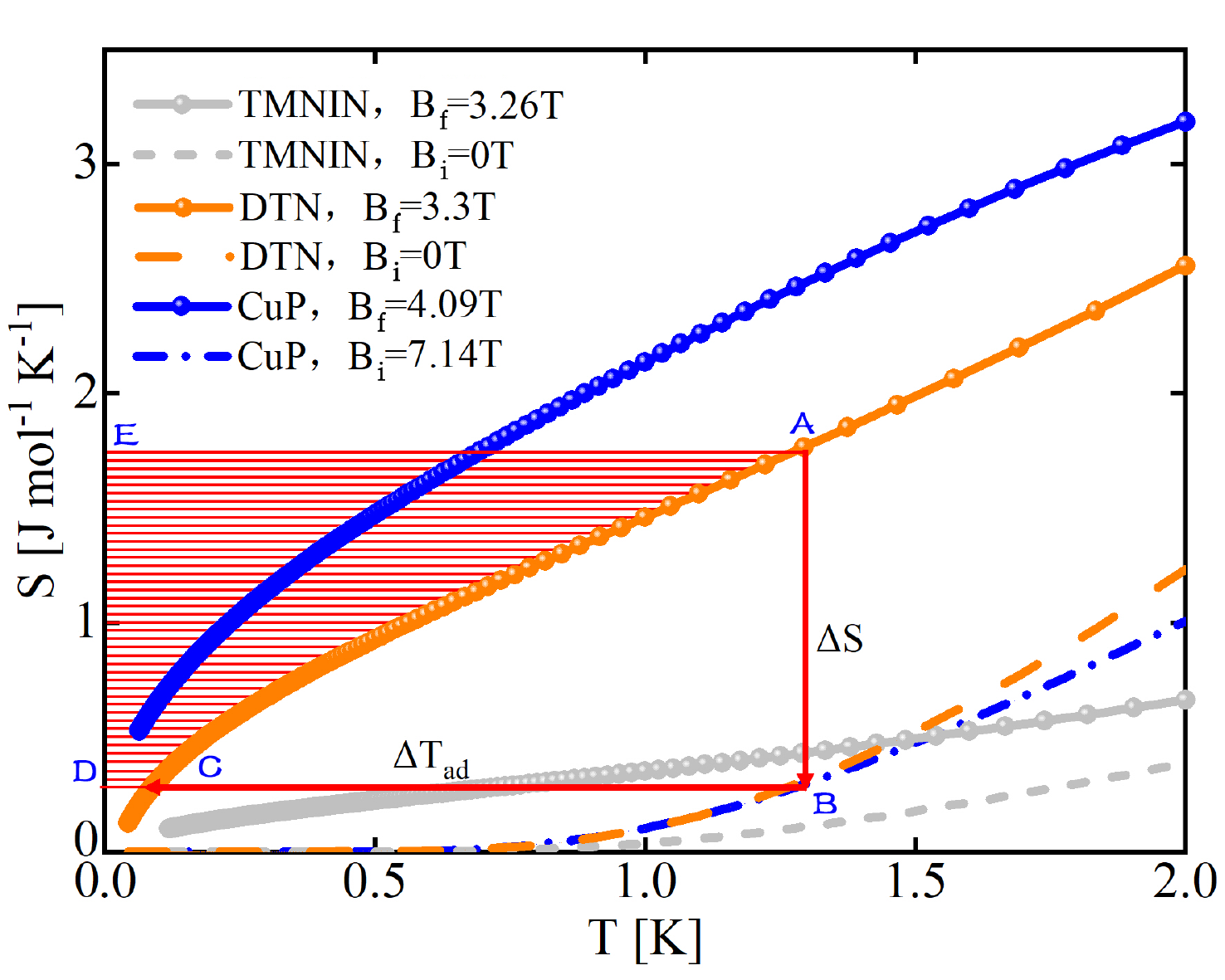}
\caption{{Simulated} thermal entropy $S(T)$ curves of three spin-chain materials, 
under different initial fields $B_i$ and final fields $B_f=B_c$. 
Entropy change $\Delta S$ (indicated by the vertical red arrow) 
and adiabatic temperature change $\Delta T_{\rm ad}$ 
(horizontal red arrow) are shown explicitly for the compound DTN.
A-B-C-A constitutes a single-shot refrigeration process, and the 
dashed regime ACDE represents the heat absorbed from the load 
in the iso-field CA process. 
} 
\label{Fig:Entropy}
\end{figure}

\section{Spin-1 Heisenberg-chain Model and Materials}
\label{Sec:ModMeth}

\subsection{Heisenberg antiferromagnetic chain and tensor renormalization group}

The $S=1$ HAFC systems with single-ion anisotropy can
be described by the Hamiltonian
\begin{equation}
\mathcal{H}=\sum_{i}\left[J \vec{S}_{i} \cdot \vec{S}_{i+1}+ 
D(S_{i}^{z})^{2}+g \mu_{\mathrm{B}} B \, S_i^z \right],
\label{Eq:S1HAFC}
\end{equation}
where $\vec{S}_{i}$ is the spin-1 operator on site $i$ 
(with $S_i^z$ its $z$ component), $J$ is the nearest-neighbor 
Heisenberg interaction, and $D$ the single-ion anisotropy parameter.
In the Zeeman term, $B$ is the external magnetic field, 
$g$ is the electronic Land\'e factor, and $\mu_{B}$ the Bohr magneton. 

In a rather wide parameter regime~\cite{Bonner1987,Golinelli1992},
the spin-1 chain model Eq.(\ref{Eq:S1HAFC}) has a non-magnetic
ground state with finite spin gap, which can be closed by applying a 
magnetic field $B$ through the field-induced quantum phase transition. 
In particular, for the case $D = 0$, the model in Eq.(\ref{Eq:S1HAFC}) 
reduces to the isotropic spin-1 HAFC model with the renowned Haldane 
gap $\Delta\approx 0.41J$~\cite{Haldane1983,Nightingale1986,White1993Spin1}, 
where the first excited state is in a spin triplet $S = 1$ sector. 
The introduction of $D$ in the spin chain can alter the size of Haldane 
gap, and drive the system into a trivial large-$D$ phase through 
a topological quantum phase transition at $D / J=0.93$~\cite{1990Effect}. 
For $D > D_{c},$ the gap reopens and scales proportional with $D$ in 
the large-$D$ limit. Besides, other off-diagonal single-ion anisotropies are 
found to be small in the two spin-1 chain compounds --- 
TMNIN and DTN --- considered in the present work, 
and we thus set them to zero in the rest of our discussion.

\begin{table*}[!htbp]
    \begin{tabular}{cccccccc}
    \hline  
    Compound & \textit{Abbr.} &$J/k_{B}$(K)    &$D/k_{B}$(K)      
    &$\Delta/k_{B}$(K)  & $g$ & $B_c$(T) &Reference\\
    \hline  
    (CH$_3$)$_4$NNi(NO$_2$)$_3$  & TMNIN    &$11.7-12$  & - &$4.1-4.5$  &$2.22-2.25$ & $3$ &\cite{Gadet1991TMNIN,Takeuchi1992TMNIN,ITO1994TMNIN,Goto2006TMNIN}\\   
    \hline
    NiCl$_2$-4SC(NH$_2$)$_2$  & DTN   &2.2    &8.9    &  3.2     & 2.26 & 2.13
    \footnote{The experimentally measured critical field is 2.2~T in DTN, 
    smaller than the theoretical 1D results of about 3.3~T, due to the
    influence of inter-chain interactions.} & \cite{Zapf2006DTN,Sun2009DTN,Sengupta2009DTN,Zherlitsyn2009DTN,
    Kohama2011DTN,Psaroudaki2012DTN}\\
    \hline 
    Ni(C$_{2}$H$_{8}$N$_{2}$)$_{2}$NO$_{2}$BF$_{4}$ & NENB    & $44.8-47.7$ & $7.5$ &$17.4$  &2.14 & $9-13$ &\cite{vcivzmar2008NENB,Orend2017NENB}\\   
    \hline 
    Ni(C$_{2}$H$_{8}$N$_{2}$)$_{2}$NO$_{2}$ClO$_{4}$ &NENP  &$46.2-48$  &$10-16$  &$13-7$  &$2.1-2.2$ & $9.8-13.4$ &\cite{Renard1988NENP,Kobayashi1992NENP}\\
    \hline 
    Ni(C$_{3}$H$_{10}$N$_{2}$)$_{2}$NO$_{2}$ClO$_{4}$ & NINO  &$50-52$   &$11.5-16$  &$10-15$  &2.2 & $8.9-11.2$ & \cite{Takeuchi1992NINO,Tao2011NINO}\\
    \hline 
    Ni(C$_{3}$H$_{10}$N$_{2}$)$_{2}$N$_{3}$ClO$_{4}$ & NINAZ  &$100-145$  & -  &$30-44$  &-- & - 
    &\cite{Takeuchi1992TMNIN,Zheludev1996NINAZ}\\
    \hline
    Y$_{2}$BaNiO$_{5}$ & -  &$250-280$  &$50-56$ &$60.5-110$  &2.16 & - &\cite{Kordonis2006Y2BaNiO5,Li2017Y2BaNiO5}\\
    \hline
    \end{tabular}
    \caption{Some common spin-1 chain compounds and their 
    microscopic Hamiltonian parameters, including the intra-chain 
    exchange $J$, uniaxial single ion anisotropy $D$, the spin 
    excitation gap $\Delta$, the Land\'e factor $g$, and the (lower) 
    critical field $B_c$. The magnets are listed in ascending order 
    in strength of the Heisenberg coupling $J$.
    }
    \label{Tab:SpinChainMat}
    \end{table*}

To simulate the finite-temperature properties and characterize
the iMCE properties of the spin-1 chain materials, we employ the 
infinite-size LTRG method~\cite{Li2011} in the bilayer form~\cite{Dong2017} 
for high-precision thermal many-body calculations down to low
temperature $T/J=0.01$. In the process of imaginary-time evolution 
(cooling), we retain up to $\chi=400$ bond states (with truncation 
error $\epsilon \lesssim 10^{-5}$) in the matrix-product thermal
density operator, which always guarantees accurate and 
converged thermodynamic results. 


\subsection{iMCE property and performance characteristics}
Driven by external magnetic fields, the spin-1 chains in the 
quantum critical regime show significantly enhanced MCE. 
To quantitatively characterize the MCE property, we calculate 
the magnetic Gr\"uneisen parameter
\begin{equation}
\Gamma_{B}=\frac{1}{T}\left(\frac{\partial T}{\partial B}\right)_{S}
=-\frac{1}{C_{B}}\left(\frac{\partial M}{\partial T}\right)_{B},
\label{Eq:Gamma}
\end{equation}
with $C_B = T  (\partial S/\partial T)_B$ the magnetic specific heat
(under an external field $B$). The Gr\"uneisn parameter $\Gamma_B$ 
measures the temperature change rate as a response to the small 
variation of the external field under an adiabatic condition. 
In the numerator of Eq.~(\ref{Eq:Gamma}) is a related differential quantity 
$$\Theta_T = (\frac{\partial S}{\partial B})_T = (\frac{\partial M}{\partial T})_B,$$ 
which measures the isothermal entropy change rate.

Correspondingly, when integrated over a given range of fields, e.g., 
$B \in [B_i, B_f]$ with $B_i$ and $B_f$ the initial and final (critical) 
fields in the iMCE process, respectively. The isothermal entropy change is
$$
\Delta S(T)=\int_{B_i}^{B_f}\left(\frac{\partial S}{\partial B}\right)_{T} \mathrm{~d} B
=\int_{B_i}^{B_f}\left(\frac{\partial M}{\partial T}\right)_{B} \mathrm{~d} B,
$$
and the adiabatic temperature change  is
$$
\Delta T_{\rm ad} = \int_{B_i}^{B_f} \Gamma_B \ T \mathrm{~d} B
= - \int_{B_i}^{B_f} \frac{T}{C_B} \left(\frac{\partial M}{\partial T}\right)_{B} \mathrm{~d} B.
$$

Besides the above MCE property characteristics, in practical applications the 
refrigeration efficiency factor $\eta$ and hold time are important quantities 
measuring their cooling performance. The efficiency factor is defined as the ratio 
$\eta = \Delta Q_{c} / \Delta Q_{m}$, 
where $\Delta Q_{c}=\int_{T_{f}}^{T_{i}} T \left(\partial S / \partial
T\right)_{B_{f}} \mathrm{~d} T$ refers to the heat absorption 
from load (indicated by the red shadow area ACDE in Fig.~\ref{Fig:Entropy}), 
and $\Delta Q_{m}=T_{i} \cdot\left[S(B_{f}, T_{i})-S(B_{i}, T_{i})\right]$ 
is the heat exchange between the material and the heat reservoir at the 
high temperature $T_i$. In multistage single-shot or continuous ADRs, 
the whole system must be optimized according to the pre-cooling 
requirements and  total weight, in which the efficiency factor is to be 
crucial~\cite{Shirron2007}. Besides, the hold time --- reflecting 
the temperature-time curve of the refrigerant in contact with constant 
heat load ---  is another important parameter for an efficient refrigeration. 
The refrigerant temperature is defined as $T_S (t) = T_S (0) + \dot Q/C_m $, 
and particularly we require the refrigerant temperature does not 
increase too rapidly (thus a long hold time), 
under a constant heat load $\dot Q$.

\subsection{Spin-1 Chain Quantum Magnets}
Distinct from the spin-1/2 chains, the spin-1 HAFC system  
has a gapped ground states~\cite{Nightingale1986,White1993Spin1} 
with symmetry-protected topological (STP) oder~\cite{Gu2009SPT}.
There has been continuous research interest in the investigation of
the spin-1 chain materials, with some prominent {examples} 
listed in Tab.~\ref{Tab:SpinChainMat}. In these compounds, 
there exists single-ion anisotropy term $D$ [cf. Eq.~(\ref{Eq:S1HAFC})] 
besides the Heisenberg interaction $J$. In this work, we are particularly 
interested in the compounds TMNIN~\cite{Gadet1991TMNIN,
Takeuchi1992TMNIN,ITO1994TMNIN,
Goto2006TMNIN} and DTN~\cite{Zapf2006DTN,Sun2009DTN,
Sengupta2009DTN,Zherlitsyn2009DTN,Kohama2011DTN,Psaroudaki2012DTN}, 
due to their very moderate critical field strength $B_{c} \leq 3$~T 
(cf. Tab.~\ref{Tab:SpinChainMat}) that is very suitable for 
magnetic refrigeration applications. Besides these two members,
there are other spin-1 chain materials in the family, including NENB~\cite{vcivzmar2008NENB,Orend2017NENB}, 
NENP~\cite{Renard1988NENP,Kobayashi1992NENP}, and
NINO~\cite{Takeuchi1992NINO,Tao2011NINO} (cf. 
Tab.~\ref{Tab:SpinChainMat}), which are very common 
frustration-free spin-1 materials that can be described 
by the Hamiltonian Eq.~(\ref{Eq:S1HAFC}). 

Despite a little dispute about the specific value of $J$ and Land\'e factor $g$, 
consensus has been reached that the compound TMNIN can be well 
described by a spin-1 HAFC with negligible single-ion anisotropy~\cite{Gadet1991TMNIN,Goto2006TMNIN}, 
through fitting the magnetic susceptibility~\cite{Gadet1991TMNIN}, 
specific heat~\cite{ITO1994TMNIN}, and magnetization curve~\cite{Takeuchi1992TMNIN}. Below, to explore the 
MCE properties of  {materials with the corresponding theoretical model}, we take the parameter set 
$J = 12$~K and $g = 2.25$ from Ref.~\cite{Gadet1991TMNIN} for TMNIN, 
and choose the set of parameters $J = 2.2$~K, $D = 8.9$~K, 
and $g = 2.26$ for DTN~\cite{Sengupta2009DTN}. 
Before discussing the iMCE properties and performances 
of the spin-1 chain refrigerants, we note that TMNIN can be 
regarded as an excellent spin-1 HAFC material that corresponds 
to a gapped ground state, while the DTN has $D/J \simeq 4$ 
and well resides in the trivial large-$D$ phase. TMNIN opens up 
the spin excitation gap due to the emergence of SPT order, 
in sharp distinction to gapless spin-1/2 materials like CuP. 
Therefore, it is interesting to compare iMCE of these two materials 
as \textit{topological} Haldane vs. {trivial} large-$D$ magnetic refrigerants.
 

\section{Inverse MCE in the spin-1 chain materials}
\label{Sec:iMCE}

Below we provide our main results of magnetothermodynamics
of two spin-1 chain magnets, the Haldane chain TMNIN and 
large-$D$ chain DTN, and compare them to the spin-1/2 HAFC 
compound CuP {(with parameters $J = 3.2$~K and 
$g = 2.33$~\cite{Crystal2007})}. We show the MCE characteristics, 
including the adiabatic temperature change $\Delta T_{\rm ad}$, 
isothermal entropy change $\Delta S$, G\"uneisen parameter 
$\Gamma_B$, and differential characterization $\Theta_T$, etc. 
The practical cooling performance like the efficiency factor $\eta$ 
and hold time are also discussed and compared in this section.


\begin{figure}[t!]
\includegraphics[angle=0,width=1\linewidth]{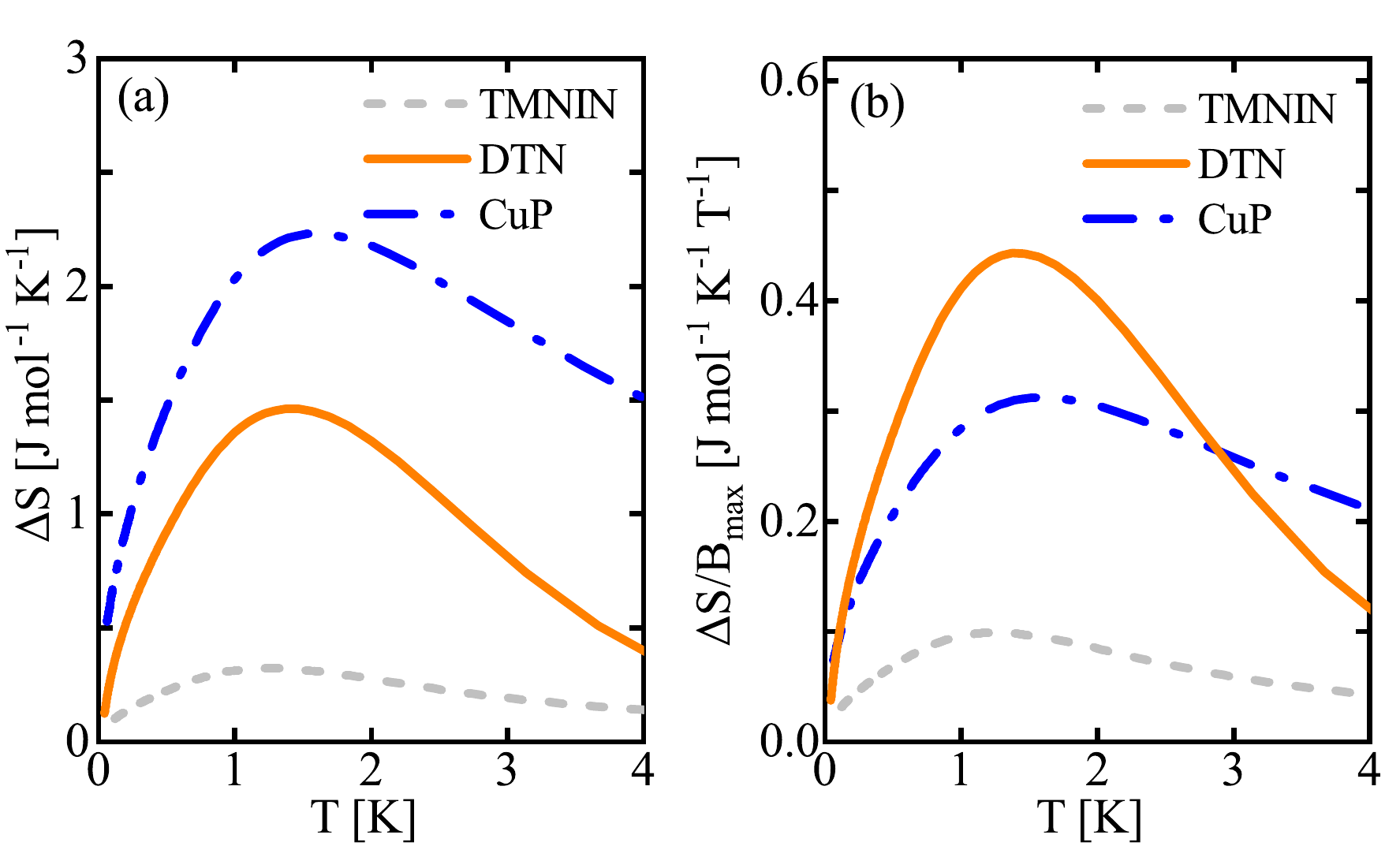}
\renewcommand{\figurename}{\textbf{Fig. }}
\caption{(a) {Simulated} molar magnetic entropy change $\Delta S$ 
as a function of temperature $T$, for three compounds,
TMNIN (gray dashed line), DTN (orange line), and CuP 
(blue dash-dot line). (b) compares the entropy change per Tesla,
$\Delta S/B_{\rm max}$, for three compounds.}
\label{Fig:DeltaS}
\end{figure}

\begin{figure*}[t!]
\includegraphics[angle=0,width=1\linewidth]{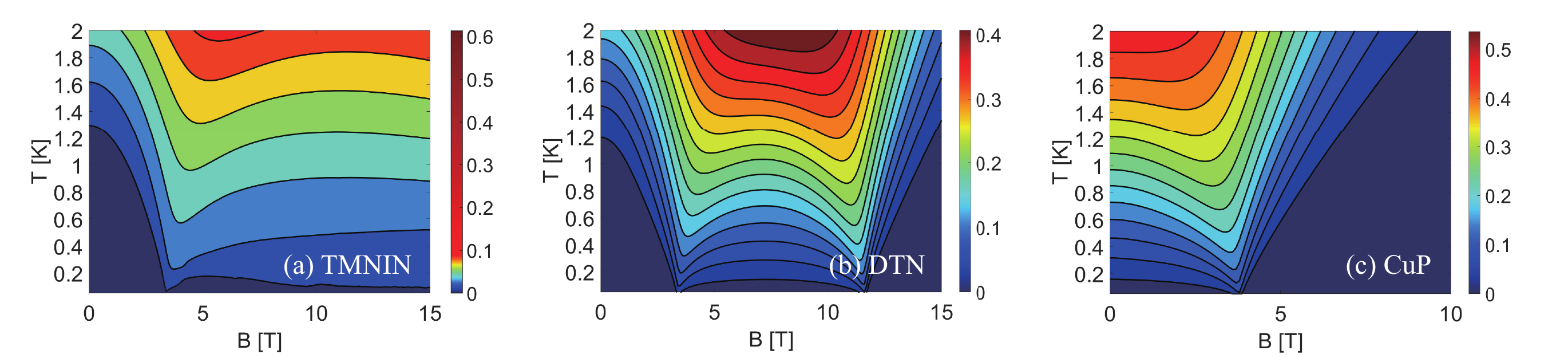}
\renewcommand{\figurename}{\textbf{Fig. }}
\caption{Simulated isentropic contour plots of three spin-chain 
compounds, including the spin-1 materials (a) TMNIN and (b) DTN, 
and (c) the spin-1/2 chain CuP. {There are two QCPs at lower 
($B_c$) and upper ($B_{s}$) critical fields in panels 
(a) and (b), where $B_c= 3.26$~T (and $B_s \simeq {32}$~T 
not shown)~\cite{2007Magnetization} for TMNIN and 
$B_c= 3.3$~\rm{T}, $B_s= 11.7 $~T} for DTN. 
Between $B_c$ and $B_s$ there exists a continuous 
Tomonaga-Luttinger liquid (TLL) regime with relatively flat 
isentropic lines at low temperature.
}
\label{Fig:Isentrope}
\end{figure*}

\begin{figure}[t!]
\includegraphics[angle=0,width=1\linewidth]{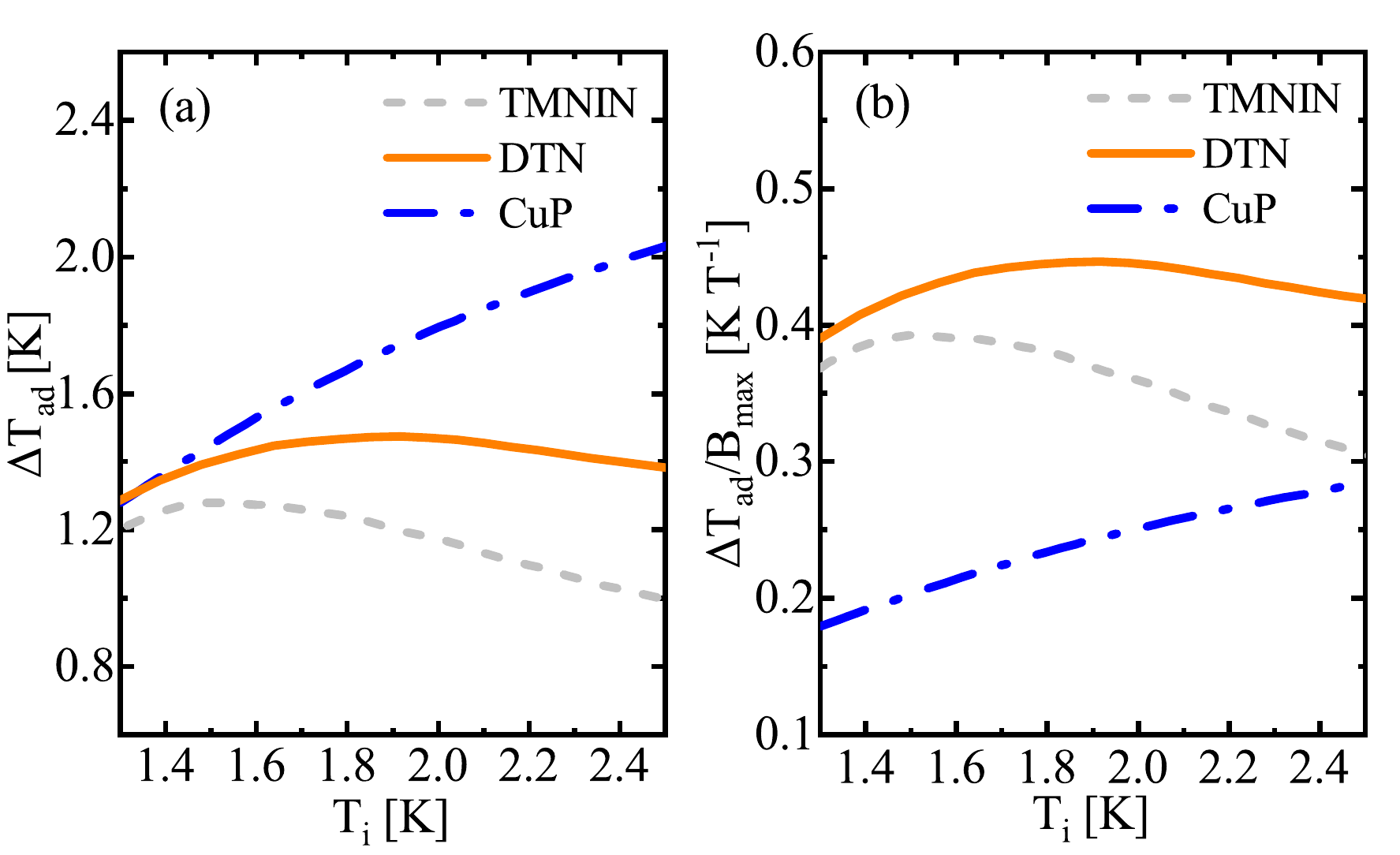}
\renewcommand{\figurename}{\textbf{Fig. }}
\caption{(a) The adiabatic temperature change $\Delta T_{\rm ad}$
and (b) temperature change per Tesla, $\Delta T_{\rm ad}/B_{\rm max}$ 
of three spin-chain compounds considered in this work.
}
\label{Fig:DeltaTad}
\end{figure}

\subsection{Entropy curves and isothermal entropy change $\Delta S$}
Magnetic fields can tune the spin states of the system and induce 
significant entropy change that can then be transferred into cooling effects.
In Fig.~\ref{Fig:Entropy}, we show the entropy curves at two 
magnetic fields --- the initial field $B_i$ and the final field $B_f$. 
As we are considering the iMCE process, they are set as 
$B_i=0, B_f = 3.26$~T (TMNIN) and $B_i=0, B_f = 3.3$~T (DTN)
for the two spin-1 compounds. It should be noted that in the iMCE process
the largest required field (here $B_f$) is right the critical field value ($B_c$). 
This is in sharp contrast to the MCE process of spin-1/2 material CuP, 
where $B_f=4.09$~T (also at the QCP) and the largest field in the cooling 
procedure is instead $B_i = 7.14$~T, much greater than that required 
in the compounds TMNIN and DTN. Such significant reduction of the 
maximal magnetic fields is important for the implementation of the 
quantum magnetic refrigeration in, say, the space applications. 

From the thermal entropy curves in Fig.~\ref{Fig:Entropy},
we find the Haldane magnet TMNIN has a rather small isothermal 
entropy change $\Delta S$, clearly less than 1 J mol$^{-1}$  K$^{-1}$ 
regardless of the working temperature (cf. Fig.~\ref{Fig:DeltaS}). 
However, the large-$D$ magnet DTN is found to conduct quite 
prominent entropy change, as seen in Figs.~\ref{Fig:Entropy} 
and \ref{Fig:DeltaS}, comparable to that of the spin-1/2 compound CuP.
As DTN only requires a maximal field ($B_{\rm max}=B_f = 3.3$~T), 
half of that for CuP, we find DTN has the highest ``efficiency", 
i.e., entropy change per Tesla, over the other two compounds 
as shown in Fig.~\ref{Fig:DeltaS}(b).

\subsection{Isentropes and adiabatic temperature change $\Delta T_{\rm ad}$}
In the iMCE cooling procedure, the compounds are firstly magnetized 
along the isothermal line AB (red arrowed line in Fig.~\ref{Fig:Entropy}). 
A larger $\Delta S$ means the greater cooling capacity, which, 
in the adiabatic process (indicated by the horizontal line BC) 
is translated into a large temperature change $\Delta T_{\rm ad}$. 
When the magnet reaches its lowest temperature $T_f$ at the point C, 
we contact the refrigerant with the heat load and it starts to absorb heat 
from there. The temperature of the magnetic refrigerant gradually rises 
up along the iso-field line CA with a fixed field $B_f$.

On this basis, it becomes very meaningful to compute the isentropes
of the three compounds and determine the adiabatic temperature change 
$\Delta T_{\rm ad}$ from there. In Fig.~\ref{Fig:Isentrope}(a,b), 
we show isentropes of TMNIN and DTN, where the criticality-enhanced 
iMCE can be clearly observed. When the magnetic field increases 
from zero ($B_i=0$) to critical field ($B_f  \ge B_c \simeq 3$~T 
for both spin-1 compounds), we find the temperature decreases 
monotonically, e.g., from about 2~K to about 1~K (for TMNIN) 
and 500~mK (DTN). The $\Delta T_{\rm ad}$ results are collected 
and shown in Fig.~\ref{Fig:DeltaTad}, which are found significant 
in the course of increasing fields for both TMNIN and DTN, 
in a wide range of initial temperatures $T_i$ shown up to 2.5~K.

In the isentropes of DTN (and also TMNIN) in Fig.~\ref{Fig:Isentrope}, 
we can recognize two dips in low-$T$ isentropic lines, 
which correspond to the two field-driven QCP. 
The lower-field one (at about 3.3~T for DTN) with
strong iMCE occurring due to the closure of spin gap, 
while the one at higher field (about 11.7~T for DTN) is the saturation 
transition where the spins become polarized. Different from the spin-1 chains, 
the results of spin-1/2 CuP in Fig.~\ref{Fig:Isentrope}(c) show only 
one saturation QCP at a field of about $B_c\simeq4$~T. 
If we increase the fields from zero to $B_c$, the iMCE in CuP 
is apparently weak; while the MCE in CuP is significant when 
decreasing from a large $B_i$ higher than 7~T [cf. Figs.~\ref{Fig:Isentrope}(c) 
and \ref{Fig:DeltaTad}(a)]. Therefore, to make a fair comparison,
we compute the temperature change $\Delta T_{\rm ad}$ per Tesla,
i.e., $\Delta T_{\rm ad}/B_{\rm max}$, in Fig.~\ref{Fig:DeltaTad}(b).
From the same (high) initial temperatures $T_i \lesssim 2.5$~K,
we find that the DTN actually has the largest ratio, and TMNIN is 
also more efficient than CuP. 

 \begin{figure*}[t!]
 \includegraphics[angle=0,width=0.8\linewidth]{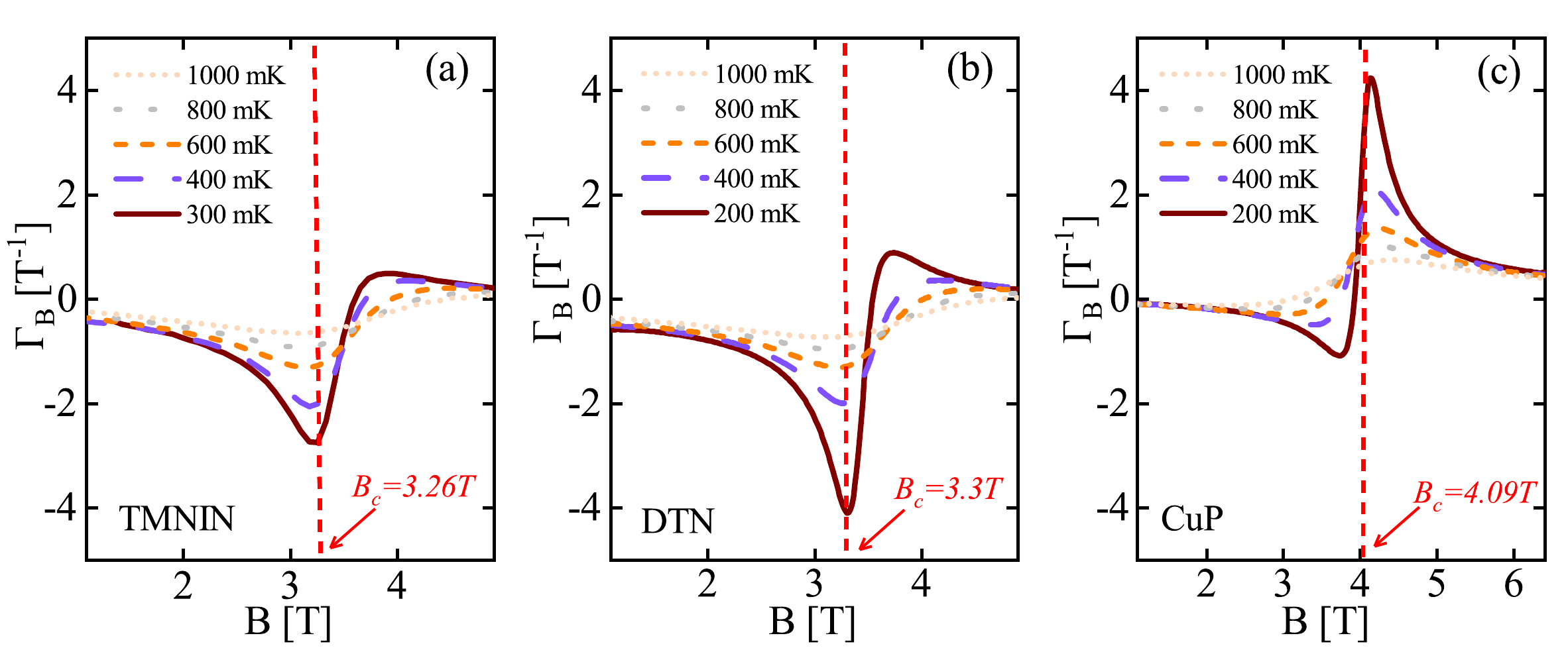}
 \renewcommand{\figurename}{\textbf{Fig. }}
 \caption{The {simulated} Gr\"uneisen parameter $\Gamma_B$ of three materials, 
 i.e., (a) DTN, (b) CuP, and (c) TMNIN, are shown at various temperatures 
from 1000~mK down to 200~mK. The red vertical dashed 
line indicates the critical magnetic fields of the three spin-chain materials.}
\label{Fig:GP}
\end{figure*}
	
 \begin{figure*}[t!]
 \includegraphics[angle=0,width=.8\linewidth]{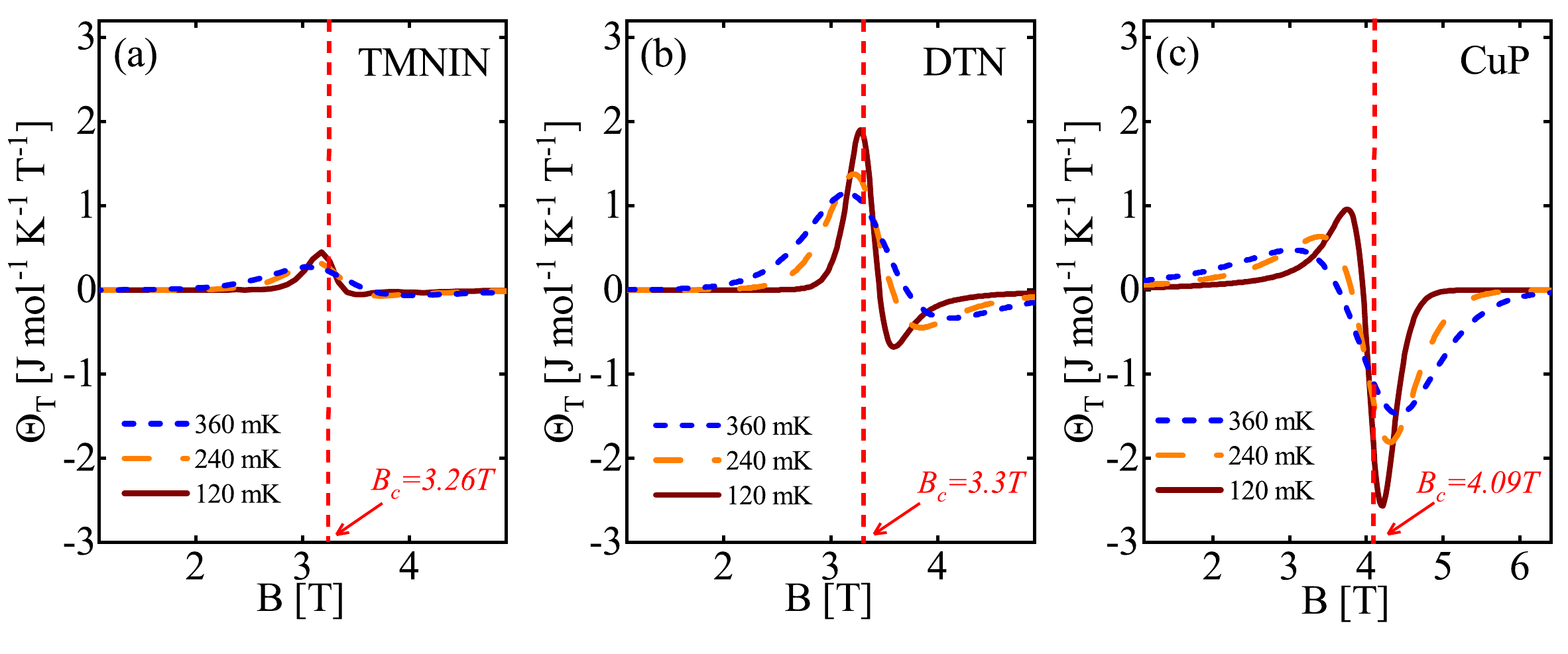}
 \renewcommand{\figurename}{\textbf{Fig. }}
 \caption{The {simulated} entropy change rate $\Theta_T$ of three materials, 
 (a) DTN, (b) CuP, and (c) TMNIN, are shown at various temperatures,
 from 360~mK down to 120~mK.}
 \label{Fig:Theta}
 \end{figure*}

\subsection{G\"uneisen parameter $\Gamma_B$ and differential iMCE $\Theta_T$}
Above we have computed the isothermal entropy change $\Delta S$ and 
adiabatic temperature change $\Delta T_{\rm ad}$ of two spin-1 materials, 
and find that the absolute and per-Tesla values are both important for 
characterizing iMCE properties. Matter of fact, to compare the MCE properties
more faithfully, by getting ride of the influences of different field ranges, 
we exploit the differential characterizations including the Gr\"uneisen 
parameter $\Gamma_B$ and differential entropy change $\Theta_T$. 
With these we are able to compare the MCE properties point by point 
at each magnetic field.

The Gr\"uneisen ratios $\Gamma_B$ of three spin-chain materials are 
compared in Fig.~\ref{Fig:GP}. In all three cases, we find pronounced 
peaks in $\Gamma_{B}$ that change its sign abruptly near the 
QCP~\cite{Zhu2003,Garst2005}, revealing the quantum criticality-enhanced  
MCE. The hight of the $\Gamma$ peak represents the adiabatic 
temperature change rate under an infinitesimal field change, 
which increases as $T$ lowers and diverges as $T \to 0$~\cite{Zhu2003}.
Again, we see that the $\Gamma_B$ peaks of TMNIN are much weaker than
those of DTN and CuP at the same temperature. The positive peaks 
and negative dips represent respectively the MCE and iMCE 
in the materials, which are quite different for the spin-1 DTN 
and spin-1/2 CuP materials. For DTN, $\Gamma_B$ is negative (iMCE) 
in the small-field side and positive (MCE) on the other, with the 
negative dip much more pronounced as compared to the positive 
peak, i.e., the iMCE in DTN is much stronger than MCE. 
On the contrary, CuP exhibits exactly the reversed behaviors, i.e., 
with a pronounced MCE peak to the right of the critical field while 
a very weak iMCE peak on the left. {This can be understood
as the TLL phase appears at different sides of the QCP in the 
two compounds CuP and DTN.}

The results of entropy change rate $\Theta_T$ are shown in Fig.~\ref{Fig:Theta},
which presents the clear positive peaks (iMCE) and negative dip (MCE) 
around the critical fields at low temperature. Similar as the observations 
in $\Gamma_B$, TMNIN again have only rather weak 
peaks(dips) in $\Theta_T$, while DTN has a pronounced 
peak with height in similar magnitude to the dip in CuP curve 
[Fig.~\ref{Fig:Theta}(c)], showing strong iMCE. Both $\Gamma_B$ 
and $\Theta_T$ display characteristic divergent behaviors close to the 
QCP and change signs as the field crosses the critical point, 
indicating the pronounced refrigeration effects through adiabatic 
demagnetization (MCE) and magnetization (iMCE).

\subsection{iMCE performance: Efficiency factor $\eta$ and hold time}
In practical applications, the refrigeration efficiency factor $\eta$ 
and the hold time are of significance to maintain a 
sustainable and high-efficiency refrigeration procedure.
The efficiency factor $\eta$ is the ratio between heat absorbed 
$Q_c$ from heat load (i.e., the area of the dashed line between 
the AC and DE lines in Fig.~\ref{Fig:Entropy}) and the released 
heat $Q_m$ to the heat reservoir, i.e., $\eta = Q_c/Q_m$.
For DTN, the area of red shadow in Fig.~\ref{Fig:Entropy} 
is $\Delta Q_{c}=0.915$~J/mol, similar to that of the spin-1/2 material CuP 
with $\Delta Q_{c}=1.144$~J/mol, while requiring less than half the field. 
For the Haldane refrigerant TMNIN, we see a small heat absorption
$\Delta Q_{c} = 0.227$~J/mol, again outperformed by the large-$D$ magnet DTN.
Besides the heat absorption, we are also interested in the heat release $Q_m$
in the isothermal process (area of the rectangle ABDE),
as those heat has to be expelled to the outer environment and thus
constitutes a load for, e.g., mechanical cooling or higher-stage ADR. 
For a robust and efficient refrigeration system, 
we want the refrigerant to absorb $Q_c$ as large as possible
while, at the same time, release smaller amount of heat to the 
environment, i.e., to have a high efficiency factor $\eta$
~\cite{Shirron2007,Jahromi2019nasa}. From Fig.~\ref{Fig:Entropy}, 
we find such a rate of the compound DTN is $\eta = 47.5\%$
when working between $T_i = 1.3$~K and $T_f=0.09$~K,  
considerably higher than that of CuP ($26\%$ from the same 
initial temperature $T_i$ as reported in Refs.~\cite{Wolf2011,Lang2013}).  

\begin{figure}[t!]
\includegraphics[angle=0,width=0.9\linewidth]{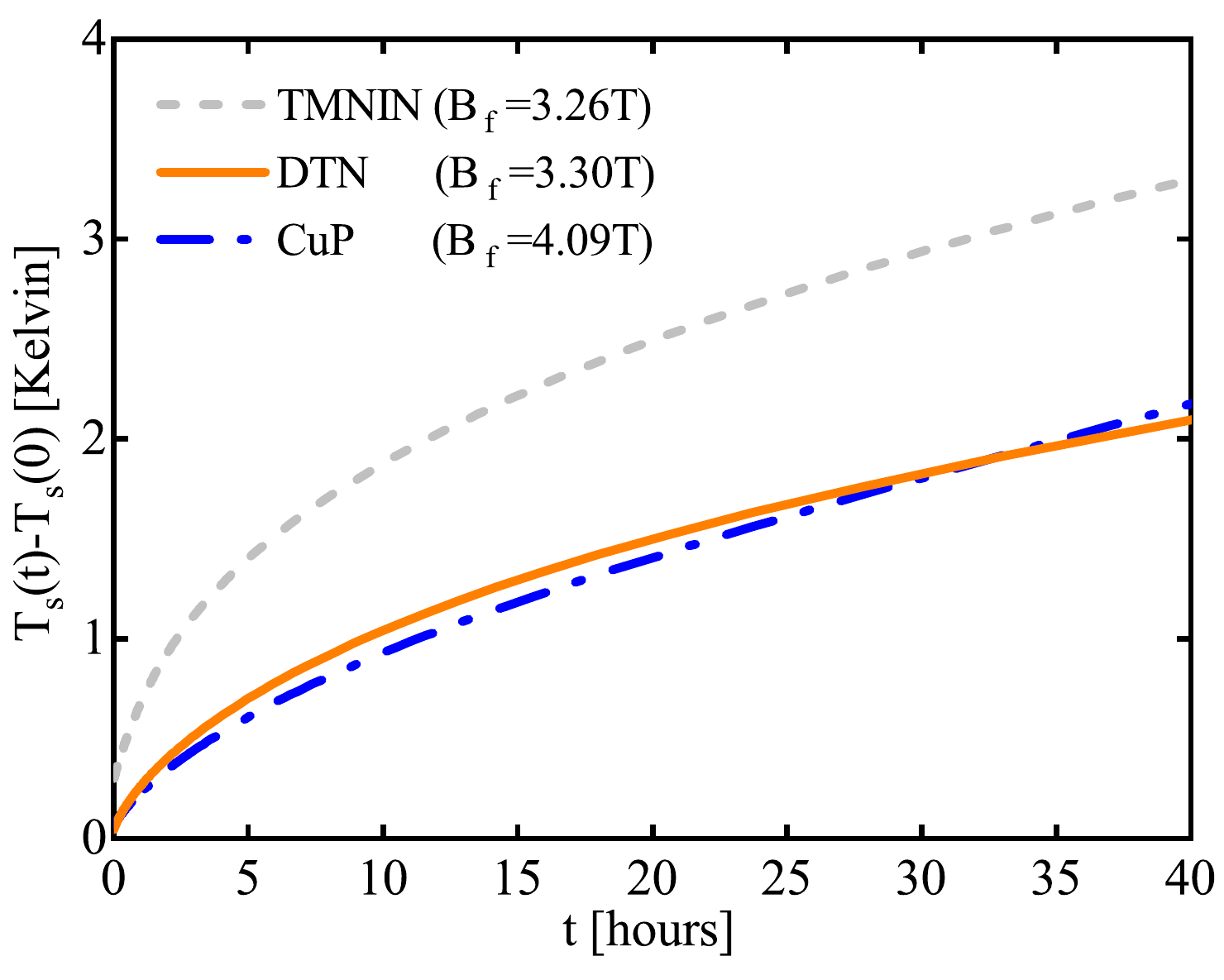}
\renewcommand{\figurename}{\textbf{Fig. }}
\caption{The sample temperature $T_s$ as a function of time $t$ at a fixed field $B=B_f$, 
calculated under a constant heat load of 5 $\mu W$. 
We consider 100 g substance of the spin-1/2 {CuP} (blue dashed line), 
TMNIN (gray dashed line), and DTN (orange solid line) in the calculations.}
\label{Fig:Holdtime}
\end{figure}

After the adiabatic demagnetization process (BC line in Fig.~\ref{Fig:Entropy}), 
the refrigerant temperature reaches the lowest value $T_f$ at the magnetic 
field $B_f = B_c$. After that, the refrigerant contacts with the load and its
temperature rises along the iso-field line CA. We want the refrigerant with 
good performance to absorb heat without warming up too rapidly. 
The simulated temperature $T_S$ at time $t$ is 
$$T_S (t) = T_S (0) + \dot Q/C_m,$$ which mainly depends on the 
magnetic specific heat $C_m$ of the refrigerant and heat load $\dot{Q}$. 
We assume that the heat is transferred at a constant rate 
$\dot{Q}=5 \, \mu W$ typical for space applications~\cite{Wolf2011}, 
and start from an initial temperature $T_{S}(0)=0.01~J/k_{\rm B}$, 
where $J$ is the spin coupling constants of the compound. The results, 
temperature $T_S (t) - T_S (0)$ versus time $t$, of three compounds are 
shown in Fig.~\ref{Fig:Holdtime}, from which we find the spin-1 DTN has 
a very similar hold time as that of the spin-1/2 CuP. The hold time of the 
latter has been shown to be quite competitive as compared to commonly
used paramagnetic salts~\cite{Wolf2011}. Overall, we find the spin-1 
magnetic refrigerant DTN is of excellent MCE performance in terms
of efficiency factor $\eta$ and hold time.

\section{Discussion and outlook}
\label{Sec:Conclusion} 
In clear distinct to the paramagnetic ADR, 
where only isolated {ions} are involved in the demagnetization cooling process, 
here in quantum magnetic refrigeration we exploit the correlation and 
entanglement of the spins and the strong quantum fluctuations
as our resource of cooling capacity, and significant temperature decrease  
can be gained in both demagnetization (MCE) and magnetization (iMCE) 
processes. Moreover, different from the conventional (classical) MCE that is 
most prominent near thermal Curie phase transition at finite temperature, 
quantum refrigeration is strongly enhanced near the quantum phase transitions 
at $T=0$. The strongly fluctuating thermal states near the quantum critical 
point prevent the constituents, spins, to freeze at low till even zero temperatures, 
resulting in large entropy change and divergent differential refrigeration 
characteristics $\Gamma_B$ and $\Theta_T$, etc, as $T$ goes to zero.

Here we have systematically investigated the iMCE near the field-induced 
quantum critical point in the spin-1 quantum magnets. As these compounds 
have a finite spin gap either due to the Haldane topological origin (in TMNIN) 
or large single-ion effects (DTN), significant iMCE have been observed when 
the spin gap is closed. In particular, the latter is found to have comparable cooling 
capacities and even better performances as compared to the criticality-enhanced 
MCE material CuP, with considerably reduced magnetic fields required.

Moreover, in the compound DTN, the field-induced quantum phase transition 
can be described as a Bose-Einstein condensation in quantum magnets, 
and has a high thermal conductivity even at very low temperature~\cite{Sun2009DTN}. 
As the typical paramagnetic salts has low thermal conductivity since the spins 
do not talk to each other in the ``gas" states, here in the spin-1 compounds 
heat can be transferred through the magnetic excitations between the coupled 
spins. This renders the spin-1 magnet DTN a very promising quantum magnetic 
refrigerant with both high cooling capacities and excellent performance.

{The compounds DTN and TMNIN are very classic spin-1 chain 
materials and their single-crystal samples have been synthesized for quite 
some time. For example, high-quality single-crystal of DTN has been used 
in the studies of Bose-Einstein condensation in quantum magnets
~\cite{Zapf2006DTN,Sengupta2009DTN} and thermal transport 
measurements~\cite{Sun2009DTN,Kohama2011DTN}. Therefore,
above we only consider the iMCE in single-crystal DTN samples,
and the fields are applied along the the single-ion $c$ axis. In case that 
only powder samples are available, or when the field tilts an angle with 
respect to the single-crystal $c$ axis, we need to consider the iMCE properties 
of DTN under a tilted field. We have also performed the calculations 
and found the field-induced QCP as well as enhanced iMCE are still 
present for a range of tilting angles (see Appendix~\ref{App:Titlted}).}

Lastly, normal and inverse MCE properties for magnets with 
high Curie temperatures have been intensively discussed 
for room- or near room-temperature refrigeration~\cite{Thorsten2005GIMCE,
Xavier2007,Naik2011,Repaka2013,Ranjan2019}, which helps 
enhance the cooling capacity and design a compact continuous 
refrigeration machinery~\cite{Zhang2007Combined}. 
Similarly, the efficient iMCE refrigerant, e.g., spin-1 DTN here, 
is important for designing a low-$T$ continuous cooling cycle where 
temperature can be decreased in both the magnetization and 
demagnetization processes. Our work fills this gap by finding DTN 
a very promising iMCE compound that provides a high-performance 
refrigerants in the ``arsenal" of spin-chain quantum materials.

 
\begin{acknowledgments}
This work was supported by the National Natural Science Foundation 
of China (Grant Nos. 11704113, 11834014, 11974036, 12074024), 
Natural Science Foundation of Hunan Province, 
China (Grant No. 2018JJ3111) and the Scientific Research 
Fund of Hunan Provincial Education Department of China 
(Grant No. 19B159).
\end{acknowledgments}


\begin{appendix}
\section{Linearized tensor renormalization group method}
\label{App:Method}
Thermodynamics of the spin-chain models and materials
can be calculated via the thermal-state tensor
renormalization group (TRG) methods. In this work, 
we employ the linearized TRG (LTRG)~\cite{Li2011,Dong2017} 
proposed by some of the authors to perform the finite-$T$ simulations.
For the spin-1 chain model Eq.~(\ref{Eq:S1HAFC}),
the Hamiltonian can be divided into odd and even parts 
through the Trotter-Suzuki decomposition~\cite{Suzuki1976}, 
and the thermal density matrix can be expressed as 
\begin{equation}
\hat{\rho}_{\beta} = e^{-\beta H} = (e^{-\tau H})^n 
= \simeq (e^{-\tau H_{\rm odd}} e^{-\tau H_{\rm even}})^n,
\end{equation}
where $n$ is a sufficiently large integer and the (small) 
imaginary time slice is $\tau = \beta/n$. In practice, 
$\tau$ is chosen as 0.05 and we iteratively project the 
imaginary evolution gates $e^{-\tau H}$ of a single Trotter 
step to the matrix-product density operator, so as to cool 
down the temperature. In the bilayer algorithm, the density 
matrix at an inverse temperature $\beta$ is obtained by
\begin{equation}
\hat{\rho}_{\beta} = \hat{\rho}_{\beta/2}^{\dag} 
\cdot \hat{\rho}_{\beta/2},
\end{equation} 
which saves the cost of calculations (by half) 
and improve considerably the accuracy by 
assuring positivity of the density matrix~\cite{Dong2017}.

\section{Quantum phase transition and magnetocaloric effects
under tilted fields}
\label{App:Titlted}

\begin{figure}[t!]
\includegraphics[angle=0,width=0.8\linewidth]{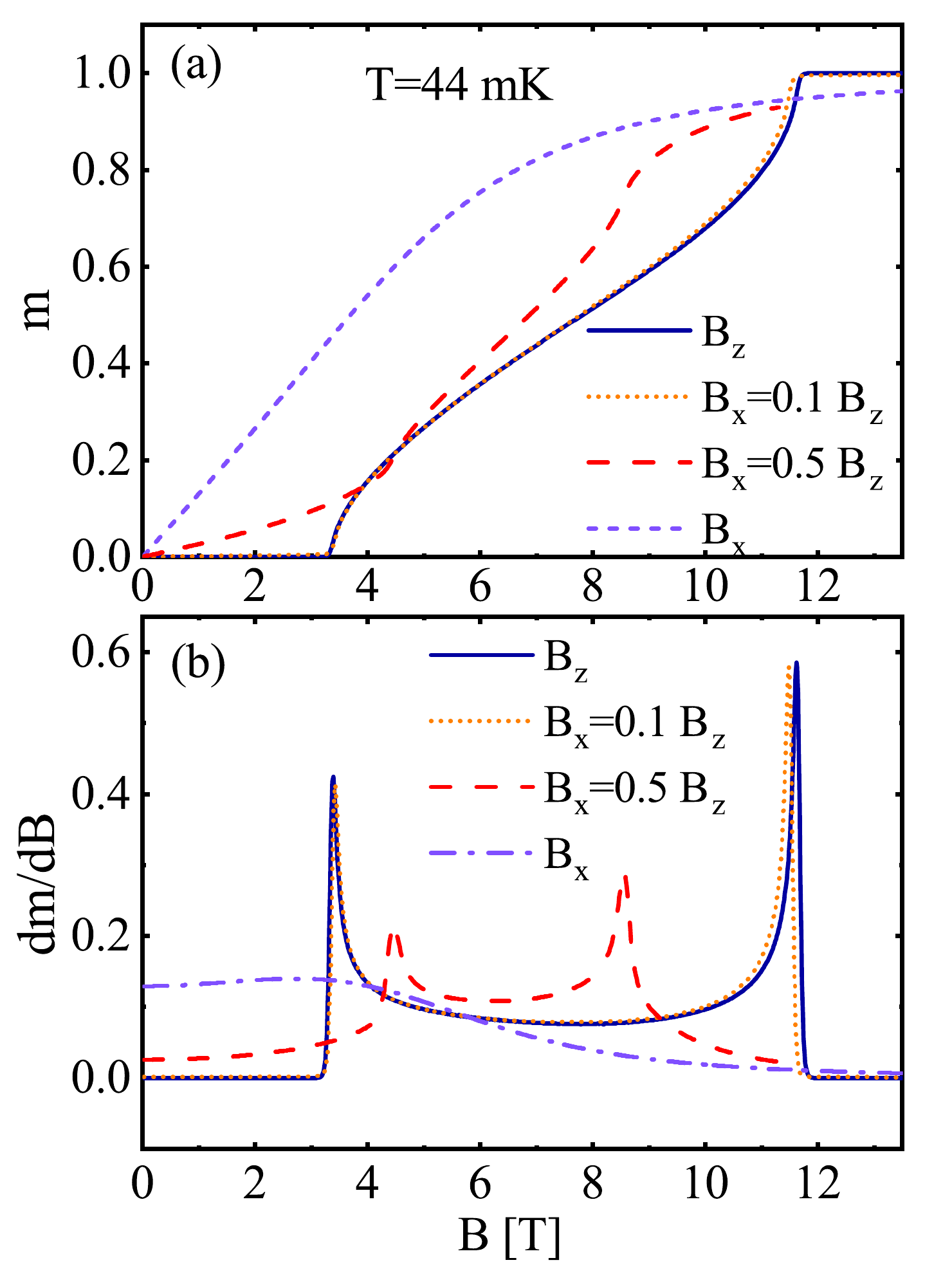}
\renewcommand{\figurename}{\textbf{Fig. }}
\caption{{(a) Low-temperature magnetization curves under magnetic 
  fields along various directions including the $c$ axis (longitudinal field $B_z$), 
  tilted directions $B_x/B_z=0.1, 0.5$, and the transverse field ($B_x$). 
  (b) shows the derivative $dm/dB$ of the three cases.}}
\label{FigR:Magn}
\end{figure}

\begin{figure}[t!]
\includegraphics[angle=0,width=1\linewidth]{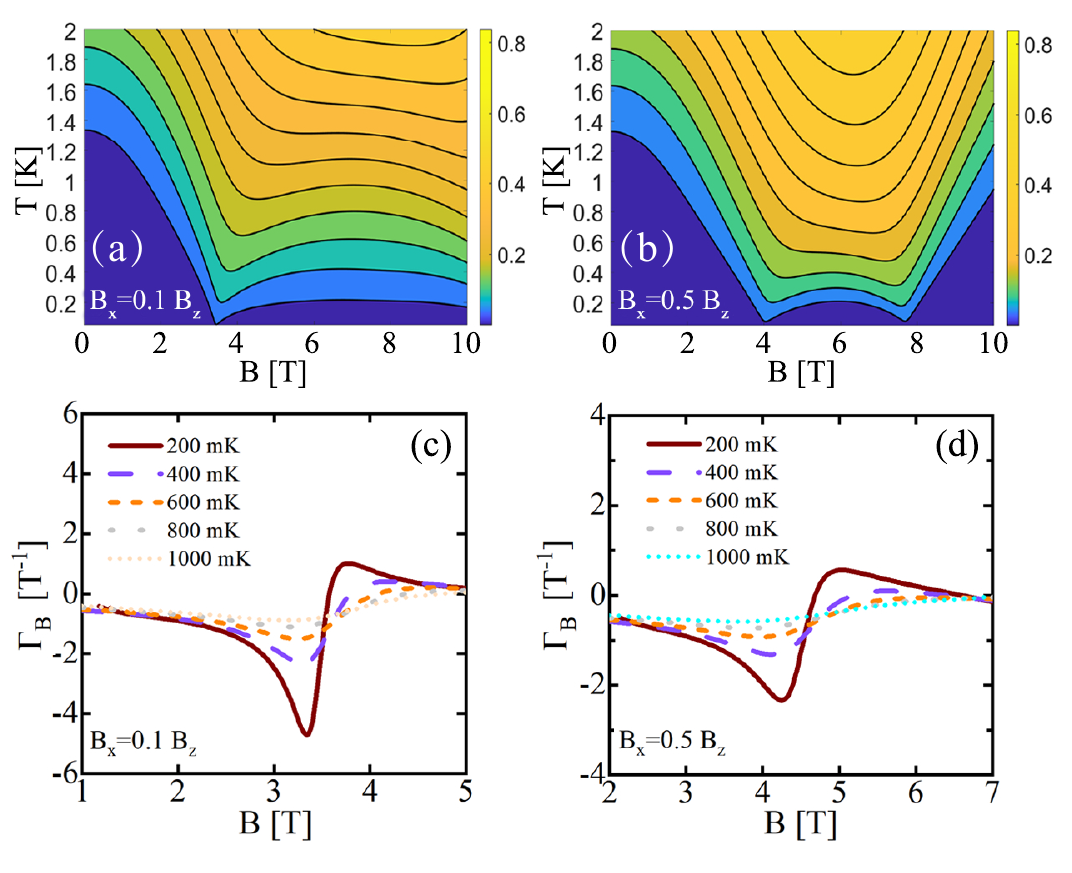}
\renewcommand{\figurename}{\textbf{Fig. }}
\caption{{(a, b) show the simulated isentropic contours and (c, d) are
the Gr\"uneisen parameters under tilted fields ($B_x/B_z=0.1$ and $0.5$). 
The lower critical fields are slightly modified in the two cases, 
and $\Gamma_B$ dip characterizing iMCE becomes less pronounced for 
$B_x=0.5 \, B_z$ as compared to that of $B_x=0.1 \, B_z$.}}
\label{FigR:GamEnt}
\end{figure}

When only powder samples are available or a small miss-alignment 
of the field direction occurs for single-crystal sample, 
the iMCE properties of DTN need to be reconsidered as the fields 
can be applied along directions other than the single-ion $c$ axis. 
In Fig.~\ref{FigR:Magn}, we show the low-$T$ magnetization process
along the tilted directions as described by the Hamiltonian
$$\mathcal{H}=\sum_{i} J \, \vec{S}_{i} \cdot \vec{S}_{i+1}+ 
D(S_{i}^{z})^{2}+ g \mu_B (B_x \, S_i^x + B_z \, S_i^z),$$
with the total field strength $B = \sqrt{B_x^2 + B_z^2}$.
In Fig.~\ref{FigR:Magn} we computed the cases $B_x = 0.1 \, B_z$ 
($\theta = 5.71^\circ$) and $B_x = 0.5 \, B_z$ ($\theta = 26.6^\circ$),
and find there still exists criticality-enhanced iMCE when tilting the 
longitudinal from the $c$ axis with an angle $\theta$. Correspondingly, 
in Fig.~\ref{FigR:GamEnt} we see clear iMCE effect near the field-induced 
QCPs, including the evident temperature decrease and clear dip in the 
Gr\"uneisen parameter $\Gamma_B$. 

On the other hand, when the field is applied perpendicular 
to the single-ion axis (i.e., $B_x$ with $\theta=90^\circ$), 
indeed we observe that the QCP is absent (Fig.~\ref{FigR:Magn}) 
and thus no pronounced iMCE can be observed. Based these calculations, 
we find the quantum criticality enhanced iMCE exists for certain tilted fields,
but not for too large tilting angles (e.g., a perpendicular $B_x$). 
Nevertheless, when the tilting angle is small (e.g., $\theta=5.71^\circ$), 
we find the iMCE rather robust and barely changed as compared to 
that along the $c$ axis. 
 
\end{appendix}

\bibliography{MCERef}

\end{document}